\documentclass[12pt]{article}
\usepackage{epsfig}
\usepackage{graphicx}
\usepackage{epstopdf}
\def \be {\begin{equation}}
\def \ee {\end{equation}}
\begin{document}

\bigskip\bigskip


\noindent\LARGE{\textbf{The emergence of quantum capacitance in epitaxial graphene$^\dag$}} 

~

\noindent\large{A. Ben Gouider Trabelsi,$^{\ast}$\textit{$^{a}$} F. V. Kusmartsev,\textit{$^{a}$} D. M. Forrester,\textit{$^{a,b}$} O. E. Kusmartseva,\textit{$^{a}$} M. B. Gaifullin,\textit{$^{a}$} P. Cropper,\textit{$^{c}$} and M.Oueslati \textit{$^{d}$}} 

~
\footnotetext{\textit{$^{a}$~Department of Physics, Loughborough University, Loughborough, United Kingdom. E-mail: A.Ben-Gouider-Trabelsi2@lboro.ac.uk}}
\footnotetext{\textit{$^{b}$~Department of Chemical Engineering, Loughborough University, Loughborough, United Kingdom}}
\footnotetext{\textit{$^{c}$~Department of Materials, Loughborough University, Loughborough, United Kingdom}}
\footnotetext{\textit{$^{d}$~Unit\'{e} des Nanomat\'{e}riaux et Photonique, Facult\'{e} des Sciences de Tunis, Universit\'{e} de Tunis El Manar Campus Universitaire, Tunisia. }}

~

\noindent\normalsize{We found an intrinsic redistribution of charge arises between epitaxial graphene, which has intrinsically n-type doping, and an undoped substrate. In particular, we studied in detail epitaxial graphene layers thermally elaborated on C-terminated $4H$-$SiC$ ($4H$-$SiC$ ($000{\bar{1}}$)). We have investigated the charge distribution in graphene-substrate systems using Raman spectroscopy. The influence of the substrate plasmons on the longitudinal optical phonons of the $SiC$ substrates has been detected. The associated charge redistribution reveals the formation of a capacitance between the graphene and the substrate. Thus, we give for the first time direct evidence that the excess negative charge in epitaxial monolayer graphene could be self-compensated by the $SiC$ substrate without initial doping. This induced a previously unseen redistribution of the charge-carrier density at the substrate-graphene interface. There a quantum capacitor appears, without resorting to any intentional external doping, as is fundamentally required for epitaxial graphene. Although we have determined the electric field existing inside the capacitor and revealed the presence of a minigap ($\approx 4.3meV$) for epitaxial graphene on $4H$-$SiC$ face terminated carbon, it remains small in comparison to that obtained for graphene on face terminated $Si$. The fundamental electronic properties found here in graphene on $SiC$ substrates may be important for developing the next generation of quantum technologies and electronic/plasmonic devices.} 

\section{Introduction}

Surface plasmons in graphene have sparked the interest of the scientific community because of their potential to provide information about the carrier density in integrated photonic data processing circuits \cite{Basov2016}. In fact, plasmons, which are collective charge excitations of the electron and hole gas in graphene, may be generated by fluctuations of the chemical potential. These fluctuations are induced by an external electromagnetic field \cite{Das2008,YatingZhang2016,Trabelsi2014,Hnida2016,Schedin2007,Ridene2012,Savelev2012,Morozov2006}. The most pronounced electron-hole fluctuations are usually created in the vicinity of the Dirac point of the electron spectrum that characterises graphene \cite{Tsvelik1985}. It is there that the average charge density vanishes. A very interesting situation arises when there is a minigap in the Dirac spectrum. On the other hand, the presence of the substrate strongly influences the plasmonic behaviour \cite{Kugel2012}. We expect to find a coupling of the plasmons with the optical phonons of the substrate. This is relevant, especially, for epitaxial graphene. The latter is sensitive to the surface quality of the SiC substrate that it is grown on \cite{Emtsev2009}. 
\begin{figure}[ht!]
\centering
  \includegraphics[width=\linewidth]{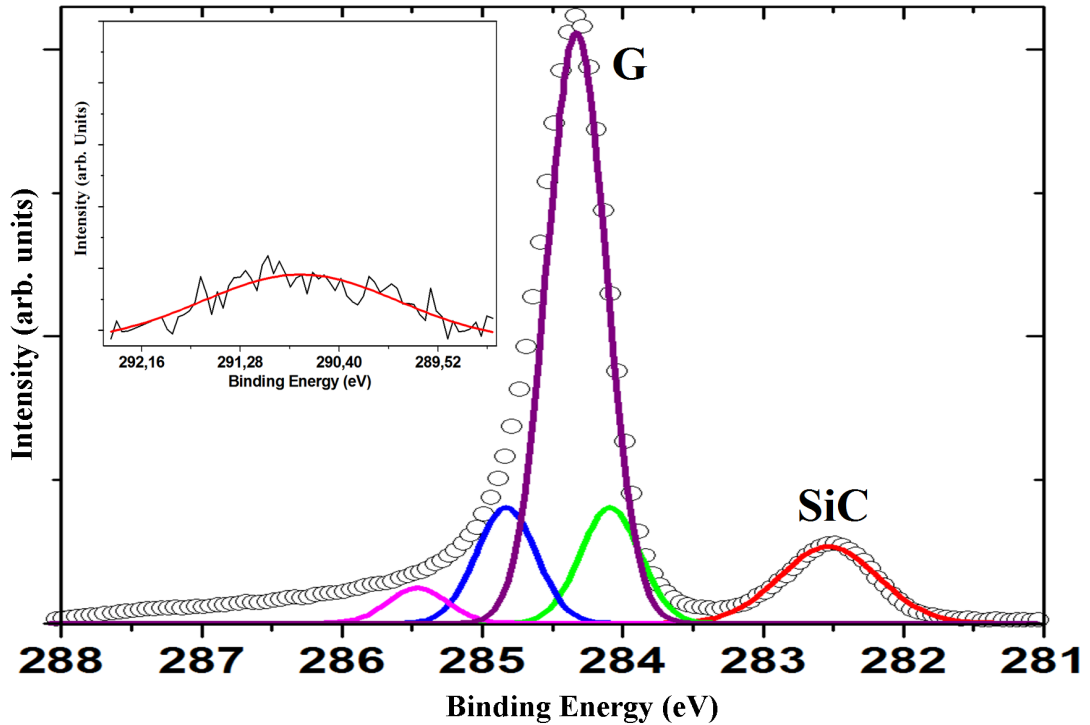}
  \caption{$XPS$ of $C$ $1s$ core level spectra of epitaxial graphene layer grown on $4H-SiC$ face terminated carbon. It shows two components at $283.6$ eV and $285.1$ eV in binding energy, attributed respectively to the $SiC$ bulk (denoted $SiC$) and the graphene layer (denoted $G$). The sharp $C$ $1s$ peak labelled $G$ and located at 285.1 eV was fitted using 4 Gaussians of $0.5$ eV width, giving a strong signature of single layer graphene. Inset: The shake-up satellite of the peak at $285.1$ eV of the face terminated carbon.}
  \label{fgr:fig1}
\end{figure} 
\begin{figure}[ht!]
\centering
  \includegraphics[width=\linewidth]{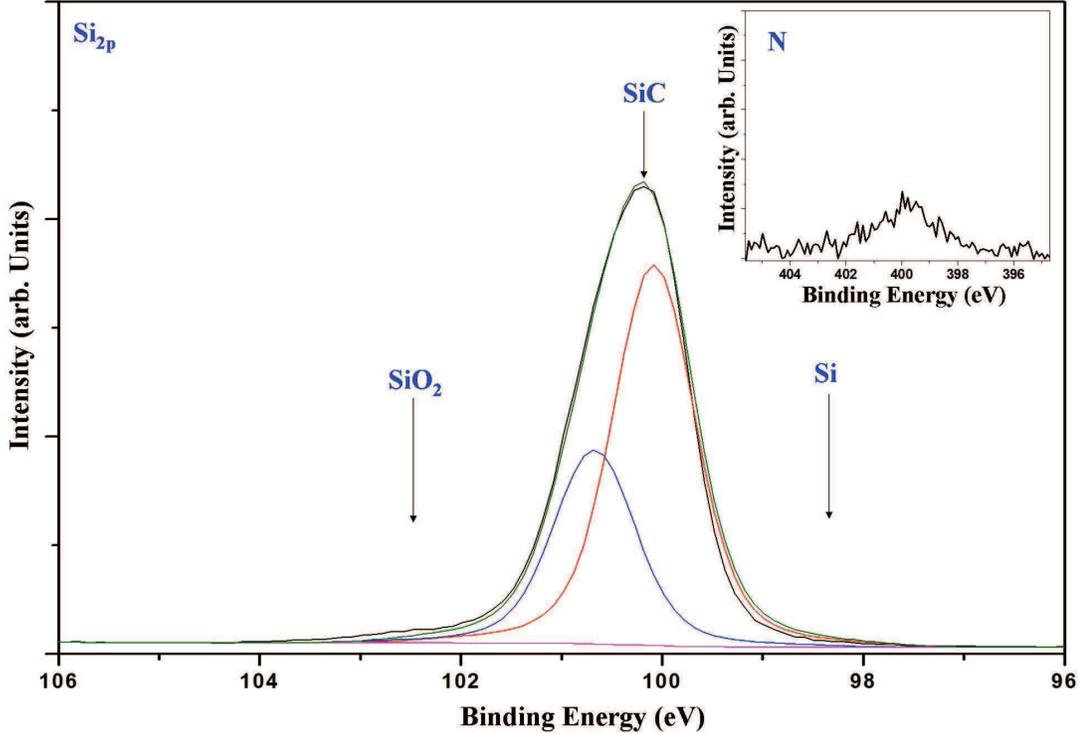}
  \caption{The $XPS$ component of $Si_{2p}$ proves the absence of any additional $Si$ or other $Si$ products (such as $SiO_2$) besides $SiC$. Inset: The $XPS$ component of Nitrogen showing its weak presence.}
  \label{fgr:fig2}
\end{figure} 
X-ray photoelectron spectroscopy ``XPS'' is a surface technique that determines the chemical composition and bonding of the surface. This kind of spectroscopy gives the average thickness of the graphene layer based on the attenuation of the substrate signal by the covering layer \cite{Emtsev2008,Ouerghi2010}. Moreover, it clearly identifies the different components forming within the graphene-substrate system. 

Raman spectroscopy has become a conventional technique for monitoring doping, defects, disorder, number of layers and phonon-plasmon coupling \cite{Ni2010,Domke2009,Casiraghi2007,Casiraghi2009,Elias2009,Nair2010}. Different Raman modes give a signature representative of an epitaxial graphene layer; mainly $D$, $G$, $G^*$, $2D$ and ($D+G$). The $G$-band is a doubly degenerate ($TO$ and $LO$) phonon mode ($E_{2g}$ symmetry) at the Brillouin zone centre, whereas the $D$-band is assigned to phonons on the $K$ point and defects \cite{Dresselhaus2005,Ferrari2006}. The $2D$ and $G^*$ bands are associated with $2TO$ and $TO~+~LA$, respectively. The ($D~+~G$) band is activated by the presence of defects in the graphene layers. Thus, the $SiC$ substrate poly-type can be identified. Furthermore, the transports properties, such as the carrier concentration, can be determined using the $LO$ phonons and their coupling with the graphene plasmons \cite{Emtsev2008,Nair2010}. 

In this work, the electronic properties of epitaxial graphene grown on $4H-SiC$ ($000{\bar{1}}$) have been investigated. A possible charge transfer due to the substrates electrostatic potential in the graphene will be discussed within. We analysed the Longitudinal-Optical Phonon-Plasmon Coupling ($LOPPC$) mode and estimated the free electron carrier concentration as a function of the graphene layer number. This method gives an approximate estimation of the charge density in comparison to that of Ref. \cite{Das2008} and \cite{Trabelsi2014}, where the areas of the electron and hole puddles, as well as any type of spatial electronic inhomogeneities in graphene, may be effectively identified. Accordingly, different electric properties, such as the electric field and the quantum capacitance, at the substrate-graphene interface were determined. We have investigated the opening of a gap at the Dirac points for face terminated carbon and we have revealed the presence of a mini-gap due to the impurity effects that affect single layer graphene. Thus, we have developed a non-invasive contactless method for measuring the charge carrier density in graphene locally, based on the Raman mapping of the $LOPPC$ mode.  

\section{Experimental details}

The confinement controlled sublimation ($CCS$) process is a commonly used method to grow graphene on the carbon terminated face of $SiC$ in a closed furnace. It was detailed by de Heer et al. in 2011 \cite{deHeer2011}. The $CCS$ is based on the silicon ($Si$) depletion from the SiC surface, with a dependency on both the local surface structure and the polarity of the face termination. At the typical growth temperatures the carbon vapour pressure is approximately $10^{−10}$ Torr, which is negligible compared to the $Si$ vapour pressure or that of the residual gases in the vacuum furnace chamber. Therefore, the process is well controlled so that for each evaporated silicon atom there remains a carbon atom left behind. Thus, the graphene monolayer is formed on the $C$-face in about $1$ min at $T = 1200^o C$ for a $SiC$ crystal that freely sublimes in vacuum. This is related to the increase of the $Si$ vapour pressure that inhibits the formation of the free carbon atoms necessary for graphene growth. Subsequently, the graphene formation temperature is shifted closer to its equilibrium (upper) value. Thus a high quality graphene monolayer is formed at $1520^o C$ for face terminated carbon. This method allows the graphitisation temperature to increase by approximately $300^o C$ compared to using a conventional ultra high vacuum method ($UHV$)  (which operates at a lower temperature, leading to the detriment of the quality of the graphene produced)\cite{deHeer2011}. 

Here, we report another approach to grow graphene on a face terminated carbon in an argon atmosphere and at lower temperature. The procedure we use is very similar to that of $CCS$ \cite{deHeer2011}, while specific details are different. The substrate we used was semi-insulating on-axis $4H-SiC$ ($000{\bar{1}}$) (C-face). The sample was exposed to hydrogen etching at $1600^o C$ in order to remove any damage due to polishing or the formation of residual oxides \cite{Srivastava2012}. The substrate was first degassed at $700^o C$ for several hours to be annealed later under a $Si$ flux at $900^o C$ to remove the native oxide. During the graphene growth process, the sample was exposed to an argon partial pressure of $P=2\times10^{-5}$ Torr and $Si$ deposition rate of one monolayer ($1~ML$)/min, while the substrate was annealed at a temperature within the range of $1200^o C$ - $1350^o C$ by electron-bombardment heating \cite{Belkhou2012}. As was found for the $CCS$ graphene growth by de Heer et al \cite{deHeer2011}, we notice that using an inert gas further decreases the growth rate since this prevents the diffusion of the evaporated silicon atoms. 
Our approach relies on the fine control of the growth mode of the graphene by precise restriction of the $Si$ sublimation rate that, in turn, regulates the release of carbon atoms. The $UHV$ chamber is equipped with a $Si$ source and Low Electron Energy Diffraction ($LEED$). The graphene layer number is evidenced by $XPS$ experiments carried out on a Kratos analytical system using an $Al$ $K\alpha$ mono-chromatised ($1486.6$ eV) source with an overall energy resolution of $\approx 350$ meV \cite{Penuelas2009,Ruoff2009}. 
\begin{figure*}[t]
\centering
  \includegraphics[width=\linewidth]{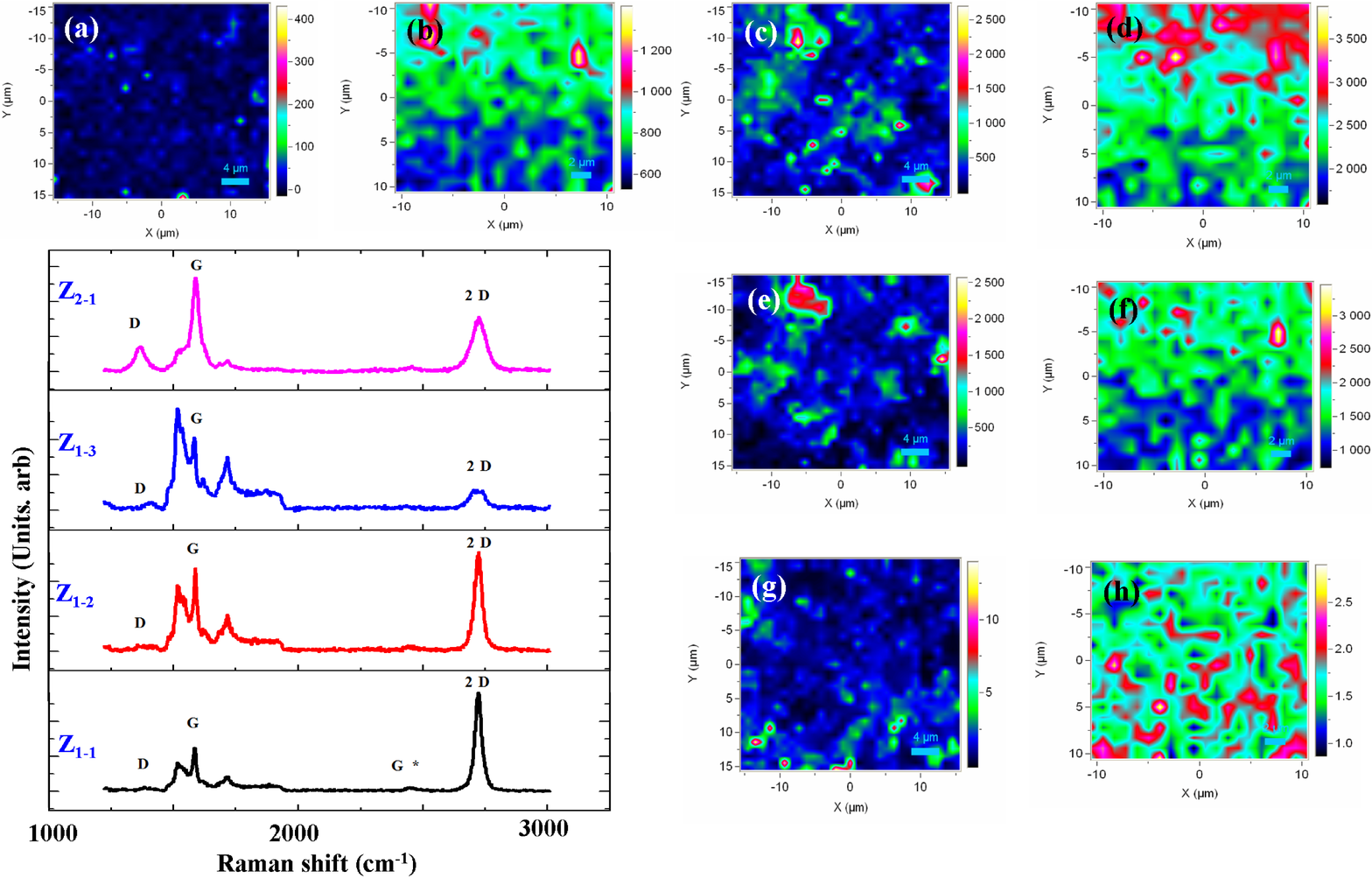}
  \caption{Raman Spectra of: single layer ($Z_{1-1}$), bilayer ($Z_{1-2}$) and graphite ($Z_{1-3}$) obtained in $Z_1$, and ($Z_{2-1}$) graphite located in $Z_2$. Raman mapping intensity of: $(a)$, $(b)$ - D band, $(c)$, $(d)$ - G band, $(e)$, $(f)$ - 2D band and $(g)$, $(h)$ - G on 2D ratio ($I_G/I_{2D}$) obtained respectively in $Z_1$ and $Z_2$.}
  \label{fgr:fig3}
\end{figure*}
Raman spectra were obtained with a high-resolution micro-Raman (Jobin Yvon HR LabRAM) spectroscope in backscattering confocal configuration. We use an $Ar^+$ laser, at the wavelength of $488$ nm, as an excitation source. The laser power was controlled at $8.5$ mW on the sample surface. We utilised a$100$X objective lens, for focusing the laser beam on the surface and collecting the scattered light, for room-temperature measurements (from different local spots forming a pixel pattern) using a grating with $600$ lines/mm to determine the graphene layer number and a grating with $1800$ lines/mm to compare the $G$ band and $LO$ phonons shifts. The spatial resolution of the image was $1 \mu m$ whilst the spectral resolution was better than $0.35 cm^{-1}$.
 
\section{Results and discussion}
\subsection{XPS measurements}
We performed $XPS$ measurements for graphene grown on $4H-SiC$ ($000{\bar{1}}$). The $C$ $1s$ core level spectra show two components at $283.6$ eV and at $285.1$ eV in binding energy. These components are attributed to the $SiC$ bulk (denoted $SiC$) and the graphene layer (denoted $G$), respectively \cite{Ni2008,Chafai2001,Heeg2013}(see, Fig. \ref{fgr:fig1}). The sharp $C$ $1s$ peak, labelled $G$ and located at $285.1$ eV, indicates the presence of $sp^2$ hybridised $C-C$ bonds. The $C_{1s}$ peak was fitted using $4$ Gaussians of $0.5$ eV width, giving a strong signature of single layer graphene that is found in compliance with techniques in the literature \cite{Mathieu2011}. The procedures used are well-known for identifying monolayer epitaxial graphene on face terminated carbon \cite{Mathieu2011} (see, Fig. \ref{fgr:fig1}). Another signature of graphitic carbon is a weak intensity peak at approximately $291$ eV. It is known as a shake-up satellite of the peak at $285.1$ eV (see the inset of Fig. \ref{fgr:fig1}). The shake-up satellite is a well-established characteristic of the photoemission process in aromatic and graphitic systems \cite{Biedermann2009}. In addition, we investigated the silicon (Si) and Nitrogen (N) flux effect on the graphene layer. 

The $XPS$ component of $Si_{2p}$, located at $100$ eV, proves the absence of any additional $Si$ or other $Si$ products (such as $SiO_2$) besides $SiC$.The $XPS$ measurements of nitrogen show a small peak at $401$ eV (see, Fig. \ref{fgr:fig2}). We could not clearly distinguish the two components $N_{1p}$ and $N_{1s}$ as a signature of the nitrogen implantation. This is due to the weak intensity found for nitrogen. The $N_{1p}$  ($397.4 $eV) is the energy range associated with pyridinic nitrogen sites in graphene \cite{Wang2013}. But, it could also be associated with other $sp^3$ and $sp^2$ $C-N$ bonds in carbon nitride films. The $N_{1s}$ is related to the $N-SiC$ bonds. Particularly, nitrogen in carbon sites bonded to $Si$ atoms at the $SiC$ interface. The binding energy of the $N_{1s}$ signal depends on the spatial repartition of the nitrogen extra charge versus the nitrogen concentration. Thus a random distribution of nitrogen may exist across the sample surface. Significant variations of the binding energies of $N_{1s}$ (between $400$ eV and $402.7$ eV) were observed, as reported in the literature, with many values comprised in this interval \cite{Joucken2015}. The variation of the $N_{1s}$ binding energy is partly due to its dependence on the local nitrogen concentration \cite{Joucken2015}. However, the wide prevalence of nitrogen in the system would be associated with a large number of defects in the graphene layer \cite{Wang2013}. A low amount of defects are confirmed below with Raman spectroscopy measurements in various areas of the samples. On other hand, our mini gap is smaller than one would find as resulting from nitrogen implantation at the graphene - $SiC $interface ($< 0.7$ eV) \cite{Wang2013}. Thus, we can not rule out the nitrogen contribution. Nitrogen could contribute to the minigap opening found here. Nevertheless, all these factors indicate that the break of symmetry remains the main factor assigned to the opening of the minigap.  
\begin{figure}[t]
\centering
  \includegraphics[width=\linewidth]{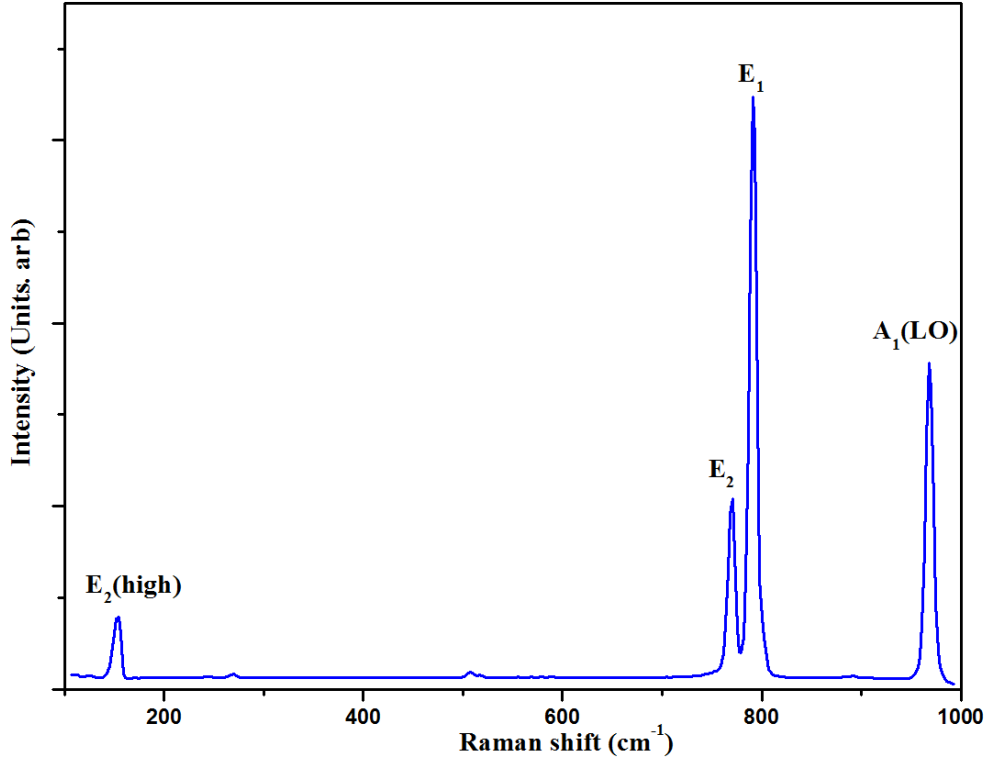}
  \caption{Raman spectrum obtained at the frequency range [$1000 - 3000 cm^{-1}$] signature of $4H$-$SiC$ ($000{\bar{1}}$) substrate.}
  \label{fgr:fig4}
\end{figure}
\subsection{Raman spectroscopy}
Careful analysis of local Raman spectra in a multitude of areas across the sample surface characterise the degree of homogeneity and the number of graphene layers, $n$. Raman mapping was carried out with a $0.5 \mu m$  step in zones $Z_1$ and $Z_2$ in turn. During mapping acquisition the laser beam focusing was checked at each point using an auto focusing adjustment. Figure \ref{fgr:fig3} shows the local Raman mappings intensity at the $D$ band frequency ($\omega_D$), the $G$ band frequency ($\omega_G$) and the $2D$ band frequency ($\omega_{2D}$), correspondingly in $Z_1$ and $Z_2$. The Raman mapping of the $D$ band displays weak intensity across the graphene sample surface in $Z_1$, eliminating the possibility of the presence of nitrogen doping effects, while it slightly increases in $Z_2$ (see Fig. \ref{fgr:fig3} $(a)$ and $(b)$). This is related to the graphite layers interaction.The Raman mapping intensity ratio of the $G$ and $2D$ bands $I_G/I_{2D}$ show similar behaviour (see, Fig, \ref{fgr:fig3}  $(g)$ and $(h)$). We have determined the number of graphene layers in $Z_{1,2}$ according to well-known procedures concerning the intensity ratio of the $G$ and $2D$ bands ($I_G/I_{2D}$) \cite{Zhan2011,CunLi2011}. The $Z_{1}$ is mainly covered by single ($n=1$) and bilayer graphene ($n=2$). Nevertheless, small graphite flakes were located. $Z_2$ is $80\%$ covered with graphite and small bilayer ($n=2$) flakes. The determination of the graphene layer number is given below. We have performed a Raman study in two frequency ranges {$(I) = [100 - 1000 cm^{-1}]$ and $(II) = [1000 - 3000 cm^{-1}]$}. The range $(I)$ corresponds to the first order Raman modes of the $SiC$ substrate, while the range $(II)$ is assigned to the second order spectral bands of the SiC substrate and to both the first and second order Raman modes of graphene. Here, we limited our study to the second range of frequencies. Figure \ref{fgr:fig3} shows the typical Raman spectra of various graphene layers located in $Z_1$ $(Z_{1-1}$, $Z_{1-2}$, and $Z_{1-3})$ and $Z_2$ $(Z_{2-1})$. Numerous second order Raman modes of $4H$-$SiC$ also appeared in the frequency range $[1479-1905 cm^{-1}]$ \cite{Yoon2009}. We have identified all the graphene peaks $D$, $G$, $2D$, $G^*$ and ($D+G$) (see, Fig. \ref{fgr:fig3}). Previous works have identified the layer number $n$ of epitaxial graphene grown on $4H$-$SiC$ using the integrated intensities ratio of the $2D$ and $G$ bands ``$I_G/I_{2D}$'' \cite{Calizo2009,Smet2008,Pinczuk2007}. For a ratio less than $0.5$, we have a single layer of graphene.  If the intensity ratio is in the range of $[0.5 - 1]$ a bilayer exists. Finally, when it is greater than $1.8$ multi-layers of graphene emerge ($n > 5$). The Raman mapping intensity ratio of the $G$ and $2D$ bands $I_G/I_{2D}$ in $Z_1$ and $Z_2$ are consistent with the layer number order, $n$, given by this established method (see Fig. \ref{fgr:fig3} $(g)$ and $(h)$) \cite{Calizo2009,Smet2008,Ni2008}. In our case, we associate the intensity ratio $I_G/I_{2D}$ of $0. 45$, $0.8$ and $<1.8$ to single, bilayer and graphite respectively, as appears in the related Raman mapping of $I_G/I_{2D}$ in both the investigated areas 
            
\begin{figure}[t]
\centering
  \includegraphics[width=7cm]{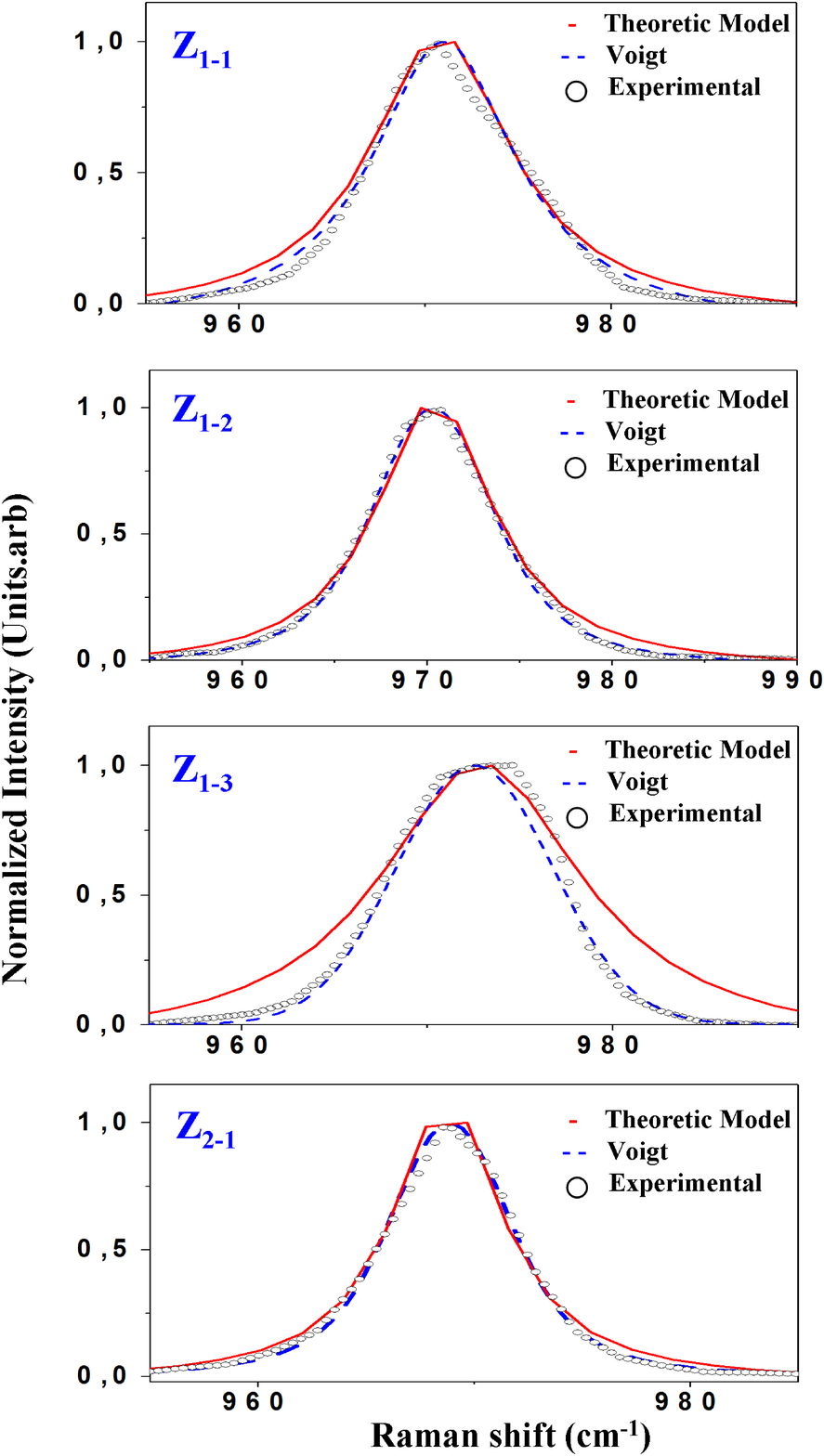}
  \caption{Variation of the line shape of the theoretical (solid line), experimental (points) and Voigt (dashed line) curves of the ``LOPPC'' modes as function of the epitaxial graphene layers for: single layer $(Z_{1-1})$, bilayer $(Z_{1-2})$ and graphite $(Z_{1-3})$ obtained in $Z_1$, and graphite located in $Z_2$ $(Z_{2-1})$.}
  \label{fgr:fig5}
\end{figure} 
\subsection{Phonon - Plasmon couplings}
We investigated the longitudinal optical phonon-plasmon interaction in the $4H$-$SiC$ substrate. We want to show that there is a strong coupling between these two modes.  We connected the energy shift from point to point on the surface to the carrier density difference in the graphene layers studied with other methods \cite{Das2008,Shestopalov2016}. LOPPC modes are bulk substrate properties. LOPPC modes have bulk substrate properties. This was well documented for $n$-type $SiC$, which has been investigated for many years \cite{NakashimaHarima1997}. Figure \ref{fgr:fig4} shows the Raman spectra of our $4H$-$SiC$ substrate, obtained in the frequency range $[100 - 1000 cm^{-1}]$, which corresponds to the major modes of the $4H$-$SiC$ substrate.  We clearly distinguish the $E_2$ (high), $E_2$, $E_1$ ($TO$) and $A_1$ ($LO$) modes observed respectively at $154$, $770$, $791$, and $967 cm^{-1}$ (see Fig. \ref{fgr:fig4}) \cite{Burton1999,Temple1973,Faugeras2008}. 

We have used a $488$ nm line of an $Ar$-ion laser as an exciting probe. Raman spectroscopy characterises the charge distribution in the whole area of the substrate for a few micrometres thickness. Particularly, we have used the high resolution confocal arrangement of the Raman spectrometer. Thus, we are able to probe to a typical depth of $1$ to $2 \mu m$. The small shift variations of the  $A_1$ ($LO$) mode, across the sample surface, are correlated with the local change in residual carrier concentration obtained using other methods \cite{Das2008}. These frequency shifts were also compared to that of the undoped $4H$-$SiC$ substrate located at $962 cm^{-1}$ \cite{Heeg2013}. So plasmonic waves, if excited, may induce charge density fluctuations both in the substrate and graphene. Therewith, they polarize the graphene and excite the $LO$ and $TO$ phonons in the substrate. The shift and the line shape of the $LOPPC$ band are analysed using a theoretical model in which the Raman intensity is given by \cite{NakashimaHarima1997,Burton1998},  
\be
I(\omega)=\frac{d^2S}{d\omega~d\Omega}\propto A(\omega)~Im(-\frac{1}{\epsilon(\omega,~q)}),
\label{Eq:equation1}
\ee
Where $A(\omega)$ and $\epsilon(\omega,~q)$ are the spectral and dielectric functions, respectively (see, the Ref. \cite{Burton1998}). $A(\omega)$ and $\epsilon(\omega,~q)$ are a function of plasma frequency, $\omega_p$, plasmon damping concentration, $\gamma_p$, damping constant of electrons, $\Gamma$, high-frequency dielectric constant, $\epsilon_\infty$, Faust-Henry constant, $C$ and phonon frequency, $\omega_T$($\omega_L$) of the $A_1$[$LO$] ($E_1$ [$TO$]) $4H$-$SiC$ mode. The theoretical curve is obtained with the following fitting parameters: the electron effective mass in SiC is $m^*=0.29*m_0$ ($m_0$ is the free electron mass) and the frequency of $TO$ ($LO$) phonons is $\omega_T=744 cm^-1$ ($\omega_L=962 cm^-1$) ), characterising an undoped $4H$-$SiC$ substrate (see, Ref. \cite{Penuelas2009}). We found a blue shift and broadening of the line width of the $LOPPC$ peak that we think is due to the increase of the phonon - plasmon interaction \cite{Pinczuk2007}. This cannot be associated with a heating effect due to the high control of our measurements setup. We reported a typical fitted $A_1$ ($LO$) spectra of single layer graphene ($Z_{1-1}$), bilayer graphene ($Z_{1-2}$) and ($Z_{1-3}$) graphite found in $Z_1$ and also graphite located in $Z_2$ ($Z_{2-1}$) (see, Fig. \ref{fgr:fig5}). The $A_1$ ($LO$) mode is not fitted properly due to its broadening. This might be  caused by the high local charge density fluctuations . The description of this broadening is at the limit of the applied model's capability. This is different from weak doping, which is characterised by the fact that the $A_1$ ($LO$) mode is well fitted (see, $Z_{2-1}$ and $Z_{1-2}$). This does not affect the determined value of the frequency. In fact, the theoretical fit generally does not properly adjust the tail of the $A_1$ ($LO$) band studied here. A Voigt fit, based on a Lorentzian-Gaussian shaped curve, provides a better fit. This has been used to determine the $A_1$ ($LO$) Raman shift position change across the sample surface, respectively in $Z_1$ and $Z_2$ (see, Fig. \ref{fgr:fig6} $(c)$ and $(e)$). The $A_1$ ($LO$) Raman mapping intensity shows similar behavior to $G$ and $2D$ bands (see, Fig. \ref{fgr:fig6} ($(a)$-$(b)$) and Fig. \ref{fgr:fig3}($(c)$-$(f)$)). The $A_1$ ($LO$) shift variation is sensitive to the local doping. The value of the shift variation is compared to the spectral line of the pure $4H$-$SiC$ ($000{\bar{1}}$) substrate i.e. undoped. This substrate was used to grow our epitaxial graphene layers. The $A_1$ ($LO$) discussed is located at $962 cm^{-1}$. A high blue shift $\delta \omega$ is observed in $Z_1$ of $11 cm^{-1}$ [$969-973 cm^{-1}$] when comparison is made to the undoped substrate (see, Fig. \ref{fgr:fig6} $(c)$). On other hand, a maximum shift of $7 cm^{-1}$ is observed in $Z_2$, [$968.4-969.4 cm^{-1}$] (see, Fig. \ref{fgr:fig6} $(e)$). Thus, the $A_1$ ($LO$) shift varies from $1$ to $4 cm^{-1}$ depending on the investigated zones. The high shift in $Z_1$ is associated with the weak layer number. Contrary to $Z_2$, a gradual dissimilarity is observed between the Raman mapping intensity and the shift variation of $A_1$ ($LO$) in $Z_1$ (see, Fig. \ref{fgr:fig6} $(a)$ and $(c)$). Moreover, the phonons shift variation between substrate and weakly doped graphite layer is consistent, as observed in $Z_2$, (see, Fig. \ref{fgr:fig6} $(b)$ and $(e)$). In fact, the presence of the highly doped graphene layer and the formation of quantum capacitance induce charging or charge redistribution in the system. In fact, the presence of the highly doped graphene layer and the formation of quantum capacitance induces charging or charge redistribution in the system. In actuality, the large electron density variation is associated with a wide shift variation of $A_1$ ($LO$). This results in dissimilarity between the shift variation of the $G$ band of the graphite and the $A_1$ ($LO$) of the substrate. Consequently, the shift variation of $G$ band of the graphite and the $A_1$ ($LO$) of substrate are not similar (see, Fig. \ref{fgr:fig6} $(c)$ and $(d)$). This is also corroborated by the behavior observed in $Z_2$ when compared to $Z_1$ (see, Fig. \ref{fgr:fig6} $(e)$ and $(f)$).     

To confirm our findings, we studied a third homogenous area of our sample, $Z_3$ of $80 \times 80 \mu m^2$ covered with single-, multi-layer graphene and graphite (see, Fig. \ref{fgr:fig7}). Here, the $I_G/I_{2D}$ ratio identifying single layer graphene is similar to the one reported in $Z_1$ (see, Fig. \ref{fgr:fig3}).The $A_1$ ($LO$) frequency varies between  $964$ and $967 cm^-1$ despite the presence of single layer graphene (see, Fig. \ref{fgr:fig7} $(f)$). This proves the sensitivity of our model to the present carrier density. But, the shift variations are not similar to those of the intensity (see, Fig. \ref{fgr:fig7} $(e)$ and $(f)$). This is regardless of the weak doping existing in this area. In fact, the $A_1$ ($LO$) shift variation is $\approx 3 cm^-1$, similar to $Z_1$. Furthermore, we noticed dissimilarity between the Raman shift variations of the $A_1$ ($LO$) substrate mode and the $G$ mode of the graphene, as also reported in data from $Z_1$ (see, Fig. \ref{fgr:fig7} $(f)$ and $(g)$). This confirms our discussions above. In fact, the large shift frequency of $A_1$ ($LO$) does not behave similarly to those of $G$ band for a wide electron density variation, regardless of whether there is a high or low charge density. Probably, this happens because in the system there exists large charge density fluctuations which result from the high screening of Coulomb forces between the substrate and graphene. Therefore, we provide a good model to investigate phonon - plasmon coupling for epitaxial graphene. Also, we present a high-quality imaging of the epitaxial graphene - $SiC$ interface system. Therefore, we can estimate the free carrier concentration $n$ from the adjusted plasma frequency  in $4H-SiC$ given by the following equation \cite{Chafai2001}: 
\begin{figure}[t]
\centering
  \includegraphics[width=\linewidth]{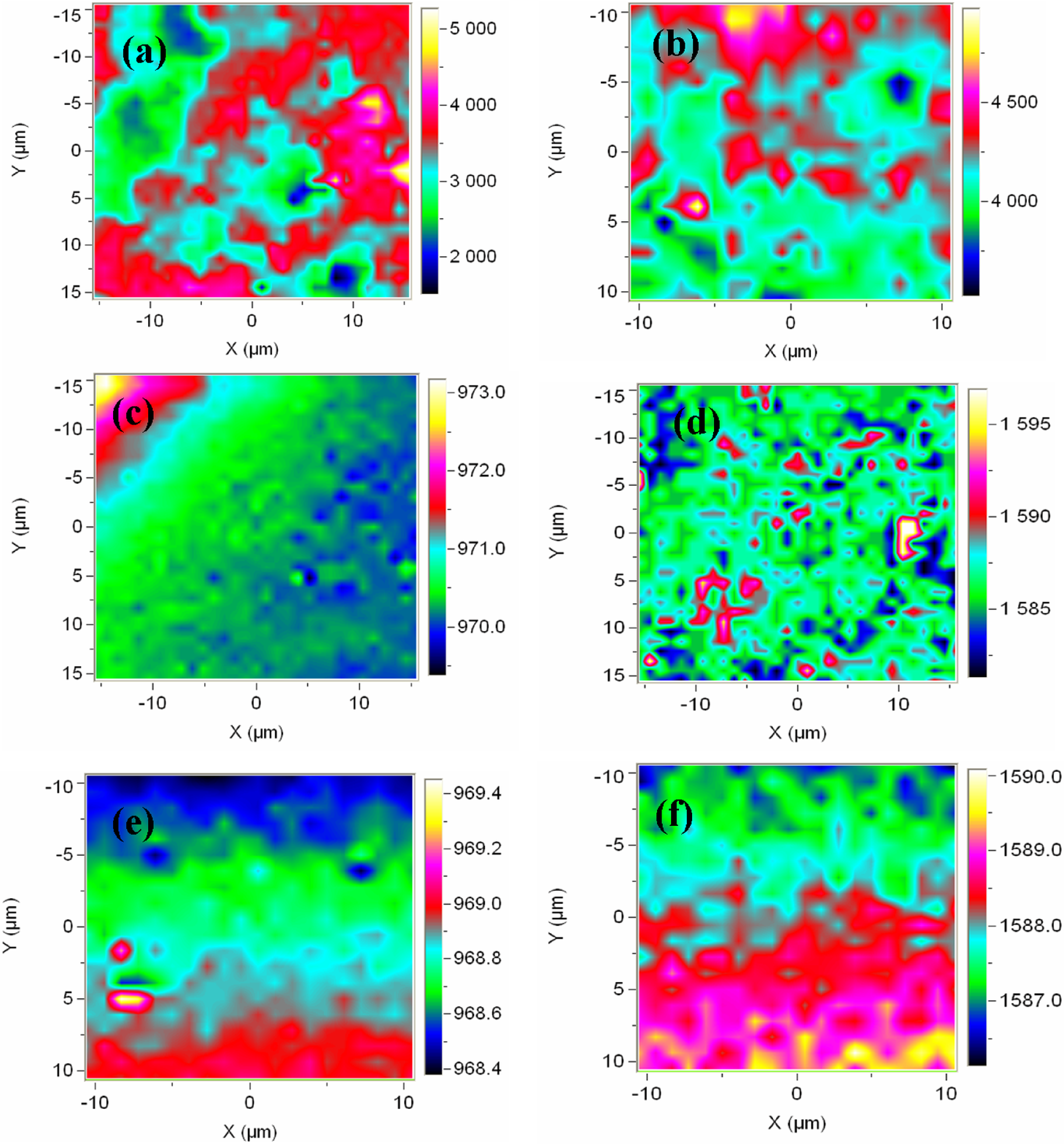}
  \caption{Raman mapping of $A_1$ ($LO$): intensity $(a)$ - in $Z_1$, $(b)$- in $Z_2$. Raman mapping of the A1 (LO) shift variation obtained in $(c)$ $Z_1$ and $(e)$ $Z_2$. Raman mapping of the G band shift variation obtained in $(d)$ $Z_1$ and $(f)$ $Z_2$.}
  \label{fgr:fig6}
\end{figure}     

\be
\omega_p^2=\frac{4~\pi~n~e^2}{\epsilon_0 ~\epsilon_{\infty}~m^*}+\frac{3}{5}(q~\nu_F)^2
\label{Eq:equation2}
\ee    
On the other hand, the plasmonic excitations in graphene, close to the Dirac point for small plasmonic momenta $q$, are given by the following expression \cite{Casiraghi2009}:
\be
\omega_{pgr}^2=\frac{2~n_{gr}~e^2}{m}~q+\frac{3}{4}(q~\nu_F)^2
\label{Eq:equation3}
\ee 
where $m$ is the electron effective mass at the bottom of the graphene band, and $n_{gr}$  is the charge carrier density in the graphene layer. An interaction appears between these two branches of plasmonic excitations given in (\ref{Eq:equation2}) and (\ref{Eq:equation3}). This results in a weak hybridisation and formation of a polariton excitation spectrum. However, we will not study such a hybridisation in the present work. Here, we have focused mainly on the phonon-plasmon interaction which is relevant for both branches and verified using the substrate properties, i.e. the $LOPPC$ model as detailed above \cite{Casiraghi2009}.
  
\subsection{Graphene self-doping}
The explored $4H$-$SiC$ substrate was not doped. One should note that our approach is limited to the Brillouin zone centre excitations or weak photon excitations with vanishing momentum, $q = 0$. This approach could not be used to fit $A_1$ ($LO$) bands for high electron concentration in the substrate, due to the contribution of the excitations out of the Brillouin zone centre and the non-parabolic form of the bands that must be taken into account \cite{Burton1998}. The electron concentrations were determined from the adjusted values of the plasma frequency, $\omega_p$, and the plasmons damping $\gamma_p$, of the common substrate for any graphene layer number (see, Table 1). 

\begin{table*}
\small
  \caption{The associated values of fittings parameters of the frequency of plasma oscillations $\omega_p$ and the constants of their damping $\gamma_p$ obtained in $Z_{1-1}$, $Z_{1-2}$, $Z_{1-3}$ and $Z_{2-1}$}
  \label{tbl:example}
  \begin{tabular*}{\textwidth}{@{\extracolsep{\fill}}lll}
    \hline
    Graphene layers located in $Z_1$ and $Z_2$  & Plasma frequency ($cm^{-1}$) & Plasmons damping ($cm^{-1})$\\
    \hline
    Single Layer $Z_{1-1}$ & 199 & 310\\
    Bilayer $Z_{1-2}$  & 185 & 248\\
    Graphite $Z_{1-3}$ & 238.5 & 454\\
		Graphite $Z_{2-1}$ & 159.99 & 129\\
    \hline
  \end{tabular*}
\end{table*}

We found a density of charge equal to $n_{Z1-1}= 2.7129\times 10^{18} cm^{-3}$, $n_{Z1-2}= 2.5214\times 10^{18} cm^{-3}$, $n_{Z1-3}=4.1904\times 10^{18} cm^{-3}$ and $n_{Z2-1}=1.8857\times 10^{18} cm^{-3}$, while its initial value on the substrate without epitaxial graphene layers was $n_{Substrate}= 3.448\times 10^{11} cm^{-3}$. These densities of charge decrease by increasing the number of graphene layers. The original electron density of $SiC$ substrate, $n_{substrate}$, is very low due to the insulator character of our substrate. To confirm our conclusions, we compared the obtained carrier density with previous studies of single layer graphene using $G$ band shift variation. We found a shift variation between [$1583-1595 cm^{-1}$] and [$1586 -1590 cm^{-1}$] in $Z_1$ and $Z_2$, respectively. This presents a blue Raman shift with respect to the $G$ line position of undoped monolayer graphene located at $1580 cm^{-1}$ \cite{Sharma2010} (see, Fig. \ref{fgr:fig6}). However, the $G$ band shift increases by increasing the layer number while in a conventional case without charge redistribution a red shift is expected \cite{Dresselhaus2005}. This could be because of a combination of doping and disorder effects \cite{Ni2010,CunLi2011}. Other epitaxial graphene work \cite{Das2008} found that the electron density equals $5\times10^{12} cm^{-2}$ for a $G$ band shift of $8 cm^{-1}$. Here, the $G$ band for typical Raman spectra, as located in $Z_1$, appears at $1585 cm^{-1}$. This shift corresponds to $3.0\times10^{12} cm^{-2}$, which confirms our findings \cite{Sharma2010}. Using our $A_1$ ($LO$) model, this shift indicates an electron density equal to $n_{gr}= 2\times 10^{12} cm^{-2}$. In reality, the charge redistribution in the substrate is very inhomogeneous. It has high density of charge near the graphene layers that decreases by going inside the bulk. Thus, the small difference between the two determined electron densities is attributable to the charge inside the bulk. Devising a model capable of determining this charge is a challenge in the study of the epitaxial graphene - substrate system. Accordingly, an agreement of the charge density obtained with the Raman shift analysis of both $A_1$ (LO) and $G$ band has been found. To investigate quantum capacitance, we limited our study to single layer graphene of which a typical spectrum was given in $Z_1$. Due to the electro-neutrality of the system, the total charge accumulated in the graphene-substrate system is equal to zero.  Thus, the total charge of the substrate should be equal to the charge in the graphene with an opposite sign. Due to the electron doping of the graphene, it will be charged negatively, while the $SiC$ substrate is charged positively.his condition gives that $n_{gr}= n_{Z1-1}~L$, where $L$ is the charged layer thickness of the substrate. We find approximately a thickness $L=0.54 \mu m$ of the charged surface layer of the SiC substrate.   This was found from examining $2857$ bilayers of $SiC$, knowing that the $Si-C$ bond length in $SiC$ crystals is $1.89$ \AA. Thus, graphene and the charged substrate layers form a ``capacitor'' system that behaves as a resonant cavity for plasmon excitations propagating along the graphene surface (see, Fig. \ref{fgr:fig8}).Additionally, it acts as a mirror, limiting and screening the penetration of the electromagnetic radiation to the uncharged volume of the $SiC$ substrate. Here, the given density of charged graphene corresponds to that of the bulk substrate. The surface charge density is determined with the $2D$ projection from the value $n_{Z1-1}$. We obtain $n_s \approx (n_{Z1-1})^{2/3} = 2 \times 10^{12} cm^{-2}$, which is similar to the charge density of the graphene as $n_{gr} \approx n_s$.  This rough estimation gives a close value, as discussed above. The substrate defects and charged impurities always create an additional electrostatic potential contribution in order to obtain such a charge redistribution. Their electrostatic potential acts as a local gate voltage that changes the charge (electron or hole) density in graphene, locally confining electrons or holes in the vicinity\cite{Forrester2015}. Therefore, the associated capacitance value is proportional to the square root of the charge density.                      

\subsection{Quantum capacitance of epitaxial graphene}
Owing to the new charge redistribution between graphene and the substrate, charged puddles arise \cite{Basov2016,Forrester2015,Yacoby2008,Yacoby2009}. These puddles are naturally created in epitaxial graphene due to the trapping potential for electrons or holes originating from the $SiC$ step terraces \cite{Forrester2015,Sutter2009,Robinson2010}. This contributed to the capacitor effect found here. The total capacitance is formed from the graphene layer, the buffer layers, and the doped layer of the $4H$-$SiC$ substrate that similarly behave as three capacitances acting in series. Therefore, the total capacitance could be defined respectively \cite{AlanWu2013}
\be
\frac{1}{C}=\frac{1}{C_Q}+\frac{1}{C_{eq~buffer~layer}}+\frac{1}{C_{eq~doped~layer~of~4H-SiC}}
\label{Eq:equation4}
\ee
where, $C_Q$ is the quantum capacitance, $C_{eq~buffer~layer}$ represents the electrostatic capacitances of the buffer layer and $C_{eq~doped~layer~of~4H-SiC}$ is the doped layer of $4H$-$SiC$. We studied graphene on face terminated carbon without a buffer layer \cite{YuAPL2013}. Thus, the total electrostatic capacitance of the system is equivalent to the summation of the capacitance of the near surface doped layer of $4H$-$SiC$. This is attributed to the inverse value estimated using the summation of the inverse capacitances defined in each $Si-C$ bilayer involved (see, Fig. \ref{fgr:fig8}).       
\begin{figure*}[t]
\centering
  \includegraphics[width=16.5cm]{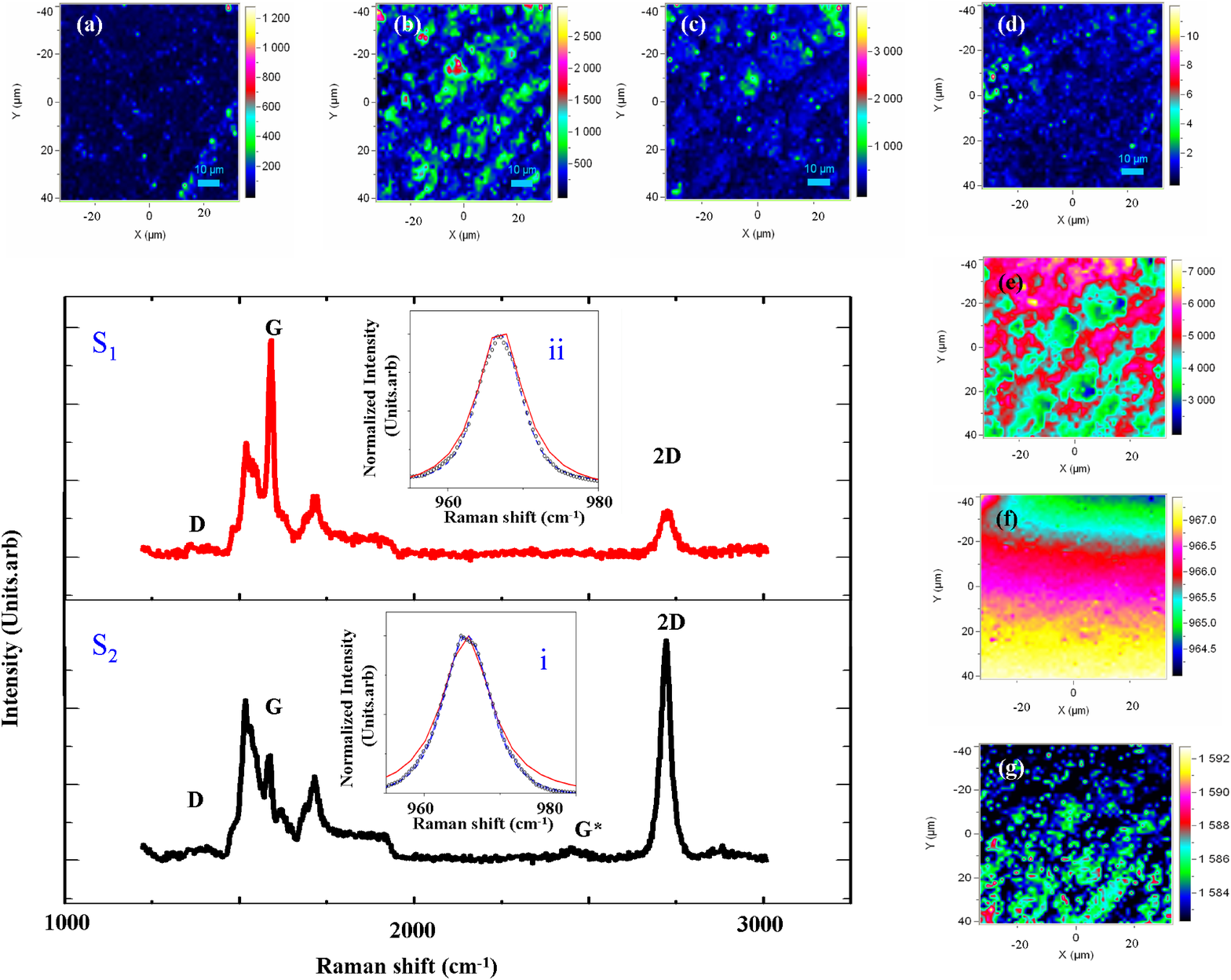}
  \caption{Raman mapping intensity of:  $(a)$- D band, $(b)$- G band, $(c)$- 2D band, $(d)$- ratio ($I_G/I_{2D}$), $(e)$-$A_1$ ($LO$) intensity. The Raman mapping of the shift variation of:  $(f)$- $A_1$ ($LO$) mode, $(g)$- G mode. The Raman spectrum of: $S_1$- single layer graphene, $S_2$- graphite, obtained in $Z_3$. Inerts: the line shape of the theoretical (solid line), experimental (points) and Voigt (dashed line) curves of the ``LOPPC'' modes associated with: i- single layer and ii- graphite obtained in $Z_3$.}
  \label{fgr:fig7}
\end{figure*} 
First, we determined the quantum capacitance $C_Q$ originating from the charged graphene layer properties. It differs from a conventional parallel plate capacitor.  The latter is formed by two plate electrodes and depends mainly on the distance between the plates, not on the charge value \cite{Trabelsi2014}.The near surface layer is self-doped in a process of the charge redistribution that occurs between graphene and the substrate in a manner similar to the construction of a Schottky barrier. To identify such a quantum capacitance, we must determine the charge density. This electron density will also depend on chemical potential, $\mu$ of the system. The density of states ($DoS$) of two-dimensional graphene is given by the expression \cite{YuPNAS2013}:
\be
D(E_F)=\frac{2\left|E_F\right|}{\pi~\hbar^2~\nu_F^2}
\label{Eq:equation5}
\ee       
We assume the electron energy is equal to the Fermi energy $E_F$, with the Planck constant $\hbar=6.58\times10^{-16}$ eV.s and the Fermi velocity in graphene, $\nu_F\approx 10^6~m s^{-1}$. Thus, the quantum capacitance can be estimated by the following equation \cite{ForresterKusmartsev2014}:
\be
C_Q=Ae^2~\frac{d n}{dE_F}=\frac{2A~e^2~\left|E_F\right|}{\pi~\hbar^2~\nu_F^2}
\label{Eq:equation6}
\ee  
where, $A$ is the surface area of the capacitor electrodes \cite{YatingZhang2016} and the electron density $n$ is related to the Fermi energy, $E_F$, via the equation:
\be
n=\int^{E_F}_{0}~D(E)dE~=\frac{g~E_F^2}{4\pi~\hbar^2~\nu_F^2}
\label{Eq:equation7}
\ee 
and $g$ is the degeneracy factor. It takes into account the double (up-down) spins and the valleys (associated with the $K$ and $K'$ points of the $BZ$) degeneracy of the Dirac spectrum of the graphene (i.e. $g = 4$). The chemical potential $\mu$, of zero applied electric field is equal to the Fermi energy, $E_F$, at low temperatures. Both depend the charge of the graphene layer.  At low temperatures, the dependence of the chemical potential (i.e., the Fermi energy $E_F$) upon the electron density $n$ is given by \cite{ForresterKusmartsev2014}:
\be
\mu=E_F=\hbar~\nu_F~\sqrt{\frac{\pi~n}{2}}
\label{Eq:equation8}
\ee   
Accordingly, the quantum capacitance is proportional to both the chemical potential $\mu$ ,counted from the Fermi energy ($\mu=E_F$),  and the degeneracy of the system \cite{YatingZhang2016}. Its contribution appears as variation arising on the top of a constant electrostatic capacitance \cite{YatingZhang2016,Appenzeller2008,Xia2009,Xia2010,Ponomarenko2010,Henriksen2010,Young2012,Feldman2012}. 
The found density of charge corresponds to the Fermi energy ($E_F = 0.14$ eV) and to a quantum capacitance per unit area ($C_Q/A=1.71\times10^{-4}~ mF~m^{-2}$). The quantum capacitance $C_Q$ is very small. Therefore, it is the dominant contribution to the total capacitance \cite{Tsvelik1985}. This is the so-called graphene quantum capacitance effect observed before in different types of graphene \cite{Giannazzo2009}. To confirm such a finding, we determined the electrostatic capacitance per unit area $C_{eq~doped~layer~of~4H-SiC}$. Such a capacitance is estimated as the summation of the $2857$ small electrostatic capacitances arising each from an individual $Si-C$ bilayer given by \cite{YatingZhang2016,ForresterKusmartsev2014}:     
\be
C_{eq}=\frac{\epsilon~S}{l_i}
\label{Eq:equation9}
\ee 
Where $\epsilon$ is the permittivity of the substrate, $S$ is the surface and $l_i$ is the $Si-C$ bond length in $SiC$ ($\approx 1.89$ \AA). This electrostatic capacitance per unit area equals $C_Q/A=1.64\times10^{-4}~ mF~m^{-2}$. In fact $C_{eq}$,  is significantly large compared to the quantum capacitance $C_Q$. Since they are connected in series, the quantum capacitance determines the total capacitance of the system. This explains the vital role of quantum capacitance effects on the total capacitance of epitaxial graphene \cite{Tsvelik1985,Xia2009}. 
\subsection{Minigap formation}
Motivated by the results mentioned above, we have examined a possible gap opening. The Dirac spectrum associated with the $K$ and $K'$ points of the $BZ$ remains invariant for the symmetry between up and down displacements. Nonetheless, the presence of a substrate underneath breaks such a symmetry owing to its additional force. Thus, the transverse lattice displacements will have different strengths applied on atoms located in up and down positions of the transverse lattice distortions. This results in a gap that could be used to estimated value of the force \cite{Casiraghi2009}. We have determined the electric field inside the total capacitor:
\be
E=\frac{e~n_{gr}}{\epsilon}
\label{Eq:equation10}
\ee
where $\epsilon=\epsilon_0~\epsilon_r$ is the dielectric constant of the $Si-C$ substrate. Taking into account the value of the dielectric constant of the $Si-C$ substrate,$\epsilon_0=10$, and the charge density $n_gr$, we found an electric field value equal to $0.43\times10^8~V~m^{-1}$. Such a field arises due to the broken symmetry between the up and down out-of-plane carbon displacements. The potential energy of a probe electron located on these two types of atoms is differentiated by the value of the electrostatic energy:  
\be
\Delta V = e~a~E
\label{Eq:equation11}
\ee
where $a=0.5$ \AA. Here $a$ is the estimation of the amplitude of the transverse lattice distortions of a suspended graphene layer \cite{Casiraghi2009}. The value of the mini-gap becomes double this potential energy difference. Substituting all the parameters into this equation we obtain a mini-gap value equal to $4.3$ meV. Its presence and the associated charge redistribution found here gives us an opportunity to conceive of new graphene-electronic devices, in which a mini-gap opening may be induced by the gate voltage. Thus, the change of the substrate properties and the associated phonon - plasmon effects were described in detail in this paper. Likewise, we revealed a possible mini-gap opening for a single layer of graphene on face terminated carbon. We also gave a detailed description of the epitaxial graphene - substrate interface. 
\begin{figure}[t]
\centering
  \includegraphics[width=8.4cm]{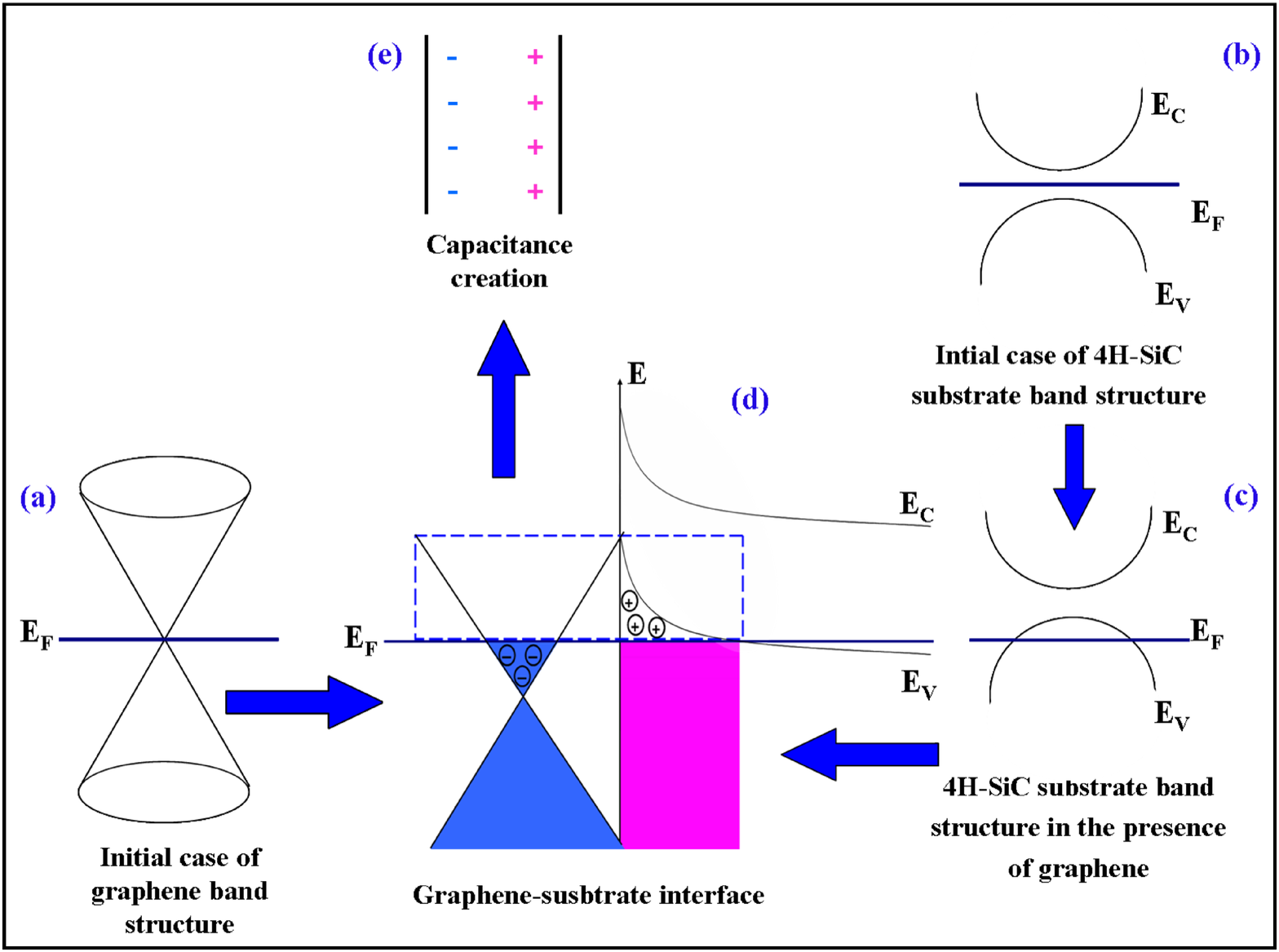}
  \caption{The capacitance formation due to mutual charge redistribution between graphene and the substrate consists of both quantum and classical capacitance: $(a)$- the initial case of the graphene band structure, $(b)$- Initial case of the $4H$-$SiC$ substrate band structure, $(c)$- 4H-SiC substrate band structure in the presence of the graphene, $(d)$- Graphene-substrate interface, and $(e)$- The capacitance creation.}
  \label{fgr:fig8}
\end{figure}     
\section{Conclusions}
In summary, we have revealed a spontaneous formation of a charged quantum capacitance in epitaxial $4H$-$SiC$ graphene. This capacitance is formed due to mutual charge redistribution between graphene and the substrate. Such capacitance consists of both quantum and classical capacitance. The formation of the quantum capacitor is associated with the spatial separation of the graphene layer from the rest of the $4H$-$SiC$ ($000{\bar{1}}$) doped substrate. The capacitor was self-charged and a mini-band gap has been determined ($\approx 4.3$ meV). Thus, our findings open a new direction for the study of self-created capacitor effects and their associated gap openings. Here, we have focused on graphene on face terminated carbon. In addition, with the use of Raman analysis several graphene characteristics, such as graphene layer numbers and disorder, have been identified. The $A_1$ ($LO$) phonon-plasmon coupled modes ``LOPPC'' of $4H$-$SiC$ substrate have been investigated. Such a coupling strongly depends on substrate doping. Thus, in this paper, we developed a non-invasive characterisation of the charge density distribution in a graphene-substrate system. Also, we gave a clear description of the epitaxial graphene - substrate interface, based on analysis of the phonon - plasmon coupling. 

\section{Acknowledgements}
We acknowledge Dr. Abdelkarim Ouerghi from CNRS ( Laboratoire de Photonique et de Nanostructures (LPN), Route de Nozay, 91460 Marcoussis, France) for providing the investigated samples.



\end{document}